# Implicit scaffolding in interactive simulations: Design strategies to support multiple educational goals

Noah S. Podolefsky*, Emily B. Moore* and Katherine K. Perkins
*University of Colorado Boulder, Boulder, CO 80309*
Email: noah.podolefsky@colorado.edu
Phone: 303-641-8217

## Abstract

We build on theoretical foundations of tool-mediated learning, tool design, and human computer interaction to develop a framework for implicit scaffolding in learning environments. Implicit scaffolding employs affordances, constraints, cueing, and feedback in order to frame and scaffold student exploration without explicit guidance, and it is a particularly useful design framework for interactive simulations in science and mathematics. A key purpose of implicit scaffolding is to support a range of educational goals including affect, process, and content. In particular, the use of implicit scaffolding creates learning environments that are productive for content learning and are able to simultaneously support the affective goals of student agency and ownership over the learning process – goals that may not be addressed in more directed learning environments. We describe how the framework is applied in the context of the *Energy Skate Park: Basics* simulation, a simulation aimed at middle school student learning of energy concepts. Interview data provides an exemplar of the process by which implicit scaffolding can support productive student exploration with a computer simulation. While we present this framework for implicit scaffolding in the context of computer simulations, the framework can be extended and adapted to apply to a range of tool-mediated learning environments.

*Authors Podolefsky and Moore contributed equally to this work, and are to be considered co-first authors.

# Introduction

*"We never educate directly, but indirectly by means of the environment. Whether we permit chance environments to do the work, or whether we design environments for the purpose makes a great difference." –John Dewey* (Democracy and Education, 1916)

Since the beginnings of formal public education, a debate has taken place over both the goals and methods of educating students. Should schooling be focused on content delivery or on student participation? Are these goals better addressed through direct instruction and rote learning, or through student-led investigation and discovery? As Dewey (1916) and others (e.g., Roschelle, 1998) have pointed out, these *either-or* questions are based on false dichotomies about the multiple purposes of education. Either-or framing of these issues has led to waxing and waning of the conventional wisdom about best practices in education. High quality research can be identified that supports either side of the debate – direct instruction and content goals (Kirschner, Sweller, & Clark, 2006) or student-centered inquiry learning and participation (Hmelo-Silver, Duncan, & Chinn, 2007). In addition, an emphasis on student agency, where students feel a heightened sense of responsibility for and control over their learning, has been advocated (Jackson, 2003; Scardamalia & Bereiter, 1991). Following Sfard (1998), we consider all of these educational goals to be valuable. The critical question is *how* to achieve these multiple objectives of education simultaneously.

Inquiry-based learning is often proposed as a means of addressing a range of content, process, and participation goals (Anderson, 2002; Minner, Levy, & Century, 2010). However, inquiry-based learning encompasses a spectrum of pedagogical approaches with varying levels and types of guidance for learners (Akaygun & Jones, 2013; Bell, Urhane, Schanze, & Ploetzner, 2010; Dewey, 1938; Reiser B. J., 2004). Activities can often involve fairly explicit guidance, such as instructions, questions, and prompts, in order to ensure students achieve certain learning goals (Lakkala, Muukkonen, & Hakkarainen, 2005; Sawyer, 2006). While using more explicit guidance can be effective for content learning, it has the potential to limit student-led exploration and achievement of participation and process goals (Bonawitz, et al., 2011; Hmelo-Silver, Duncan, & Chinn, 2007; Lin, et al., 2012). Activities can also involve very little explicit guidance - often referred to as *discovery* or *discovery-based learning*. Discovery emphasizes educational goals that more explicitly guided approaches do not, such as student agency, but its efficiency and efficacy for achieving content goals has been questioned (Kirschner, Sweller, & Clark, 2006). In this paper, we propose a framework for *implicit scaffolding* in learning environments which aims to support content, process, and participation goals while also allowing students to assume more agency and control over their own learning (Scardamalia & Bereiter, 1991).

We focus on the design of one tool used in STEM learning environments, interactive computer simulations (or *sims*). We draw on fundamental theories of learning, tool design, and human computer interaction to build the *implicit scaffolding* framework. In the framework for implicit scaffolding, scaffolding is built into the tool itself – using affordances, constraints, cueing and feedback – in such a way that students find productive inquiry paths and are supported in their learning without requiring explicit instructions. Implicit scaffolding *guides without students feeling guided*, and in doing so provides new opportunities for supporting productive inquiry while simultaneously promoting student agency.

This paper presents a theoretical framework for implicit scaffolding, demonstrates the use of this framework for design of an interactive computer simulation, and grounds the theory with a case study. The case study, consisting of an interview with a student using the simulation, serves as an exemplar of how implicit scaffolding supports multiple educational goals, illuminating student exploration patterns and their interaction with, interpretation of, and response to specific design features.

# Scaffolding in Science Learning

Scaffolding, as described by Wood, Bruner, & Ross (1976), "*consists essentially of the adult 'controlling' those elements of the task that are initially beyond the learner's capacity, thus permitting [the learner] to concentrate upon and complete only those elements that are within his range of competence.*" Scaffolding in this sense is informal, such as mother-infant interactions. Today, the definition of scaffolding has expanded to include a broad range of formal structures, including conversational devices (e.g., guiding questions), curriculum design (e.g., direct instructions for students), and features of computer software (Pea, 2004; Quintana, et al., 2004; Tabak, 2004).

The idea that learner competence is not fixed, and may be greater with appropriate scaffolding, has been influential in the development and application of scaffolding in education. Vygotsky (1978) suggested that when working with others, learners can work within a *zone of proximal development (ZPD)*, in which the learner's competence is greater than when working alone. Vygotsky identified three zones: what the learner can do on their own, can do with assistance, and cannot do even with assistance. Ideally, by working within the zone of proximal development, learners internalize knowledge and practices and their competence grows. Stemming from Vygotsky's work, scaffolding has become conceptualized as a tool to support the transition of knowledge and practices from the learner's zone of proximal development into the zone of what learners can accomplish on their own.

Scaffolding in inquiry-based learning environments is an increasingly active area of research (Lin, et al., 2012; Linn & Bat-Sheva, 2011). With the goals of increasing students' domain knowledge and their metacognitive awareness of the inquiry process, common forms of scaffolding science inquiry learning include use of written prompts (Kolodner, Owensby, & Guzdial, 2004; Reiser, et al., 2001), scripting of social interactions and argumentation (Stone, 1993), and use of representations in the form of visual models or analogies (Chiu & Lin, 2005; Podolefsky & Finkelstein, 2006).

These scaffolding methods are typically explicit, consisting of written or verbal directions to support student learning. Explicit scaffolding is commonly used within technology-enhanced environments (Akaygun & Jones, 2013; Lakkala, Muukkonen, & Hakkarainen, 2005; Linn, Davis, & Bell, Internet Environments for Science Education, 2004). While many studies support that scaffolding can improve student learning (Reiser B. J., 2004), some researchers have raised the concern that too much explicit scaffolding can result in decreased student agency (Hmelo-Silver, Duncan, & Chinn, 2007; Lin, et al., 2012; Spencer, 1999) and learning (Sweller & Levine, 1982).

*The Premise of Implicit Scaffolding*

Here we present a framework for *implicit* scaffolding. Within this framework, the scaffolding is neither written nor verbal, but is built into the design of the learning environment or the learning tool itself. By incorporating implicit scaffolding, the learning environment or tool (e.g. a sim) can support students to learn and move into their ZPD with minimal explicit guidance from a teacher or worksheet. Furthermore, implicit scaffolding can provide an inherent flexibility that can support students along varied, individualized learning trajectories, fulfilling the need for adaptability in scaffolding.

The *more knowledgeable other* described in traditional forms of scaffolding maintains influence over the student's learning in that expert knowledge and guidance is embodied in the learning environment or tool itself, and made present through its design. Sims, for instance, can embody knowledge about: the content and processes of the discipline (e.g., physics) (Latour, 2005) as well as pedagogical content knowledge (e.g., key relationships to highlight and common student challenges to address) (Shulman, 1986). In this way, learning tools that employ implicit scaffolding continue to provide students with the guidance of the more knowledgeable other, but can dramatically change the nature of how students receive and perceive the guidance.

We do not aim to minimize the teacher's role, nor do we intend for the implicit scaffolding to be the only scaffolding. Rather, by offloading some of the scaffolding responsibilities onto the learning environment or tool, the teacher then has more flexibility to adapt their role in the classroom to meet students' needs. The teacher can reduce their role as "source of knowledge" and increase their use of pedagogical practices and modes of facilitation that develop science process, reasoning, and argumentation skills, support student agency, and encourage community participation. As we will discuss, this approach can create opportunities for student engagement that are well suited to support multiple educational goals, including student agency during the learning process.

## PhET Interactive Simulations

The framework for implicit scaffolding described here emerged over the course of 10 years of research and development work by the PhET Interactive Simulations group at University of Colorado Boulder. The PhET group has focused on leveraging the unique capabilities of computer technology to create sims that support students to:

1. **Engage in scientific exploration.** Students pose their own questions, design experiments, make predictions, and use evidence to support their ideas. Students build on their prior knowledge, monitoring and reflecting on their understanding as they explore.
2. **Develop conceptual understanding.** Students develop an understanding of expert models. Students draw cause-effect relationships and coordinate across multiple representations.
3. **Make connections to everyday life.** Students connect formal science ideas to their everyday life experiences and recognize science as a tool for understanding the world.
4. **View science as accessible and enjoyable.** Students engage in authentic science practices and develop their identity as a person who uses scientific reasoning. Students demonstrate further interest in science.
5. **Take ownership of the learning experience.** Students perceive a sense of agency where they can direct their own scientific exploration.

These goals for students are well aligned with the new *Framework for K-12 Science Education* report (Quinn, Schweingruber, & Keller, 2011) and research on how students learn (Brown & Cocking, 2000). The critical challenge is to design sims that can simultaneously support all five goals – sims that engage students in scientific exploration in a way that develops their conceptual learning, increases their enjoyment, and allows students a greater sense of agency and ownership over their use of this new learning tool and over the learning experience as a whole.

The general framework and specific strategies for implicit scaffolding developed iteratively over the course of designing and testing 125 sims on diverse topics and for diverse educational levels – a process which involved over 600 individual student interviews and numerous classroom studies. Many elements of implicit scaffolding are present in our early work – where we refer to principles to support engaged exploration and implicit guidance through sim design (Adams, et al., 2008; Podolefsky, Perkins, & Adams, 2010) – as well as in studies of effective multimedia design principles identified by other researchers (Mayer, 2009).

For each new sim design, we conduct a series of individual think aloud interviews with students using the sim. Typically, students openly explore the sims, with no accompanying written or verbal guidance. With this interview methodology, we can examine how the design of the sim – including the implicit scaffolding – is engaging students, cueing and guiding productive exploration, and supporting sense-making around the learning goals. Specifically, as the student interacts with the sim, we look for: 1) usability – that the sim interface and controls are intuitive and easy to use and that the visual representations are interpreted as desired by the target student population; 2) engagement – that the student is actively exploring the sim, generating their own questions and ideas to explore and test, drawing connections, and sense-making; and 3) learning – that the student's interactions and sense-making with the sim support achievement of the learning goals. These interviews reveal a wealth of information about design features that are working well and those features that need to be redesigned or refined. Over the years, the large number of design challenges posed by the diversity of sim design topics and target age groups, combined with the insights provided by interviews and classroom observations, have continuously expanded the toolbox of approaches PhET uses to implement effective implicit scaffolding in sims. (Hensberry, Moore, Paul, Podolefsky, & Perkins, 2013; Lancaster, Moore, Parson, & Perkins, 2013)

Having developed a preliminary, tacit understanding of implicit scaffolding through empirical studies of sim use, we seek a theoretical basis on which to build a robust design framework. In what follows, we draw on existing theories and research on learning, tool design, and human computer interaction to develop the implicit scaffolding framework.

## Theoretical Foundations

In our observations of students learning with sims, we note two key findings: 1) students build understanding through exploration and by drawing on prior knowledge, and 2) the design of the sim is critical to the learning process. We therefore ground our thinking in constructivism (Piaget, 2001) and tool-mediated learning (Vygotsky, 1978). We further emphasize three key aspects of modern views of these learning theories: learning is an active process (Brown & Cocking, 2000; Mayer, 2009); students build knowledge from multiple resources (di Sessa, 1988; Smith III, di Sessa, & Roschelle, 1994; Hammer, Elby, Sherr, & Redish, 2005); and tools, such as computer simulations, play a critical role mediating learning and knowing (diSessa A. A., 2001; Hutchins, 1995; Norman D. A., 2002; Vygotsky, 1978).

With this general perspective on learning, we focus our discussion on tools – e.g., books, lab equipment, and computer simulations – and the particular aspects of tools that support learning. We begin with a broad framing of tool-mediated learning and tool design. Following this, we present the implicit scaffolding framework in detail, describing specific strategies for designing tools that provide students the opportunity to construct understanding of science concepts through an (inter)active process. The framework and the specific strategies draw on supporting literature from cognitive psychology and learning sciences, as well as our own research on student learning with computer simulations.

## Tool-Mediated Learning

The use of tools in education influences students learning and their learning process (Vygotsky, (1978)). In *Tool-mediated Learning* (Figure 1), the *learner* interacts with a *learning objective* (e.g., knowledge of conservation of energy), mediated by some tool (e.g., a computer simulation). While the learning objective is often content knowledge, it could also be scientific reasoning, enjoyment, community participation, or a combination of these (or other) goals.

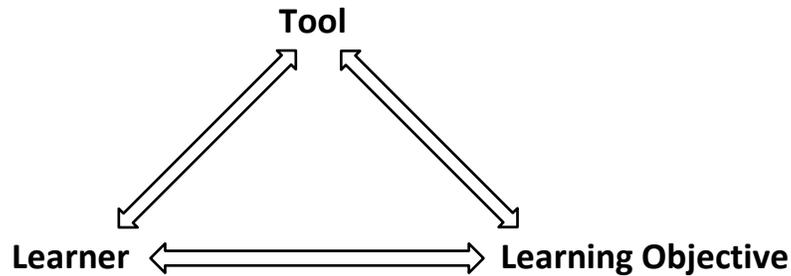

**Fig. 1** Schematic diagram of tool-mediated learning

The design of the tool impacts how learners interact with the content, participate in learning activities and construct knowledge. Different tools support different types of interaction, environments, and objectives. For instance, a textbook may be well suited to support increasing content knowledge, while shared whiteboards may be well suited to promote community participation. In selecting and designing a learning tool, it is important to attend to and appropriately leverage the unique possibilities presented by a particular genre of tools. Thus, our theoretical foundation also draws on the literature of tool design.

## Tool Design: Affordances and Constraints

To better operationalize how tools can mediate learning, we draw on a model that describes tools and their use in terms of *affordances* and *constraints* (Norman, 2002). Affordances are features of a tool that allow certain actions (Gibson, 1977). For instance, the handle of a coffee cup *affords gripping*. The cup also affords *holding liquid* (and keeping it hot). According to Norman (2002), in order for affordances to be used, they must be perceived. For instance, the coffee drinker must perceive that the handle affords gripping before it will occur to them to grip the cup this way. If this example seems trivial, it is because we live in a culture filled with coffee cups and coffee drinkers. Thus, this affordance is readily perceived because it is highly salient in our culture. One could also use the handle for hanging the cup on the wall. Use of this affordance is less common, but within certain cultural contexts it may be readily perceived. In short, it is critical to good design to make affordances of tools salient within the culture and context that they appear.

Constraints are features of a tool that restrict actions. Constraints are *productive* when the limitations they place increase the likelihood of intended usage. For instance, a typical coffee cup holds only so much liquid before spilling over. This constraint assures that the weight of the full cup can be lifted and tipped easily. In this case, perceiving the constraint will lead to the coffee drinker typically pouring an appropriate amount of coffee. (If this constraint is not perceived initially, it will become apparent relatively quickly.) Proper use of the coffee cup is fairly assured. This is the mark of a productive constraint.

The importance of attending to affordances and constraints in design cannot be understated; it can often explain the difference between good and poor designs. Effective designs use affordances and constraints to create tools where correct usage is *guaranteed by design* (Norman, 2002). Poor designs – those that lead to unintended or incorrect usage and consequently frustration – have often failed to attend to these design ideas. As described by Walsh (2007)Figure 2 (left) shows a door that needs to be *pushed* to open, yet includes a handle that affords *pulling*. Thus, it is likely that a user of this door will perceive that the handle should be pulled. The owner tried to remedy the situation by inscribing "PUSH" on the door. This type of textual direction is often a sign of poor design. Note that this door handle does afford pushing, but this affordance is not readily perceived.

In Figure 2 (right), we have edited the handle out. This door also has the affordance that it can be pushed, and cues users where it should be pushed. However, it now has the productive constraint that it *cannot* be pulled. With this design, there is no possibility that a person would mistakenly pull on the door. No instruction is needed because productive usage is guaranteed by design. Similarly, good sim design affords productive usage, but also constrains to restrict unproductive usage.

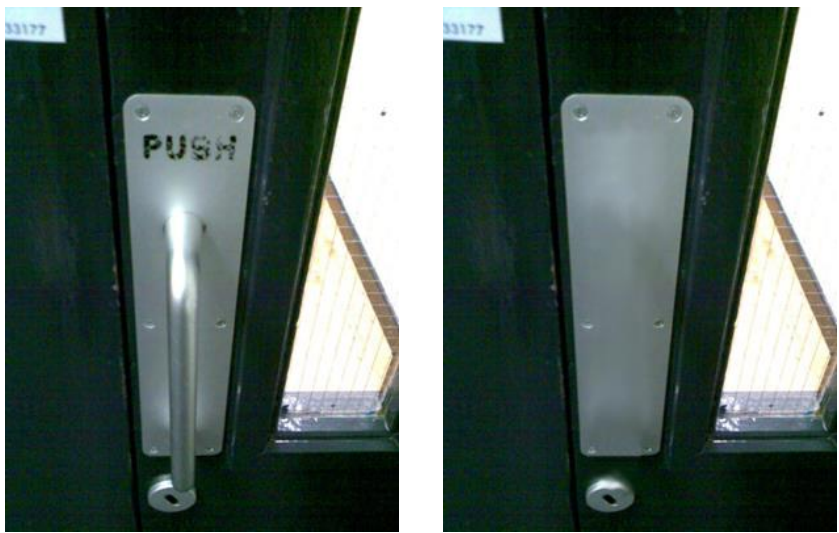

**Fig. 2** A poorly designed door (left), and a well-designed door (right). Both doors require a push. *Image on left used with permission* (Walsh, 2007)

# Framework for Implicit Scaffolding

Using these perspectives on *tool-mediated learning* and *tool design* and drawing on the findings from education research, the implicit scaffolding framework describes an approach to the design of learning environments, specifically interactive sims, that can support achievement of PhET's multiple learning objectives (listed above) simultaneously. Informed by approaches to pedagogical scaffolding, we present a set of design strategies we use for implicit scaffolding, including how they stem from the above perspectives and the broader research literature. We divide our description of implicit scaffolding strategies into four categories: Scaffolding 1) the general concept, 2) students' framing of sim use, 3) to enable sense making, and 4) for continued engagement.

*Scaffolding the General Concept*

Early in the development process for each PhET sim, the scope, sequence and interactivity of the sim is decided.

**Scope:** The scope of each sim is defined by a set of specific learning goals. By narrowing the focus of a sim from a general topic to specific learning goals, we create a tool that mediates which concepts students engage with and how students interact with those concepts to construct knowledge. (Reiser B. J., 2004; Quintana, et al., 2004) The learning goals – focused and well-defined – provide direction for further implicit scaffolding.

**Sequence:** Following traditional approaches to scaffolding (Collins, 2006), complex concepts can be parsed, layered, and sequenced through the use of tabs. Each tab in a sim provides a different environment for student exploration, focusing on a subset of specific learning goals. Within each tab, we are able to influence how students engage with the learning goals, including how the concept is presented and how affordances and constraints are used to scaffold student interaction (Norman, 2002). In some sims, successive tabs add complexity in the form of new situations, representations and/or sim tools. Sometimes successive tabs relax constraints present in earlier tabs, e.g., by allowing control over more variables simultaneously. This design approach also provides an opportunity to fade some components of the scaffolding designed into the sim as a student progresses through the tabs.

**Interactivity:** Each tab's learning goals are parameterized into interactive representations – components of the sim students can move and change. These interactive representations are designed specifically so that interaction results in immediate, dynamic feedback that supports conceptual sense making. (Lajoie, 2005; Oblinger, 2004) In deciding which representations and actions to make interactive, we must decide how to appropriately model the physical system in each sim. Each sim includes *pedagogically appropriate simplifications* – model simplifications that maintain emphasis on the targeted conceptual relationships. These simplifications are consistent with, and often based on, the types of simplifications that teachers and textbooks use when introducing material and that experts use in their own thinking about the core science ideas. Through design of the interactive representations and actions in each sim tab, we provide scaffolding for students that mediates their investigation and learning of science concepts encountered during sim use.

*Scaffolding Students' Framing of Sim Use*

*Framing* can be considered the answer to the question "what sort of activity is this?" (Hammer, Elby, Sherr, & Redish, 2005; Tannen, 1993) When students first begin interacting with a sim, we want the sim's design to encourage students to: engage in inquiry; explore and make sense of the representations, their behavior and relationships; and actively engage in the learning process. To achieve these goals, we design the opening screen to support students in framing the sim use as exploratory and focused on sense making.

**Opening Screen:** What the student sees when the sim first opens is designed to be interesting, but relatively simple to interpret. We do not want the student to start their sim use feeling overwhelmed, or to think that the sim is overly complex (Podolefsky, Adams, Lancaster, & Perkins, 2010; Norman D. A., 2011).

**Initial Interaction:** Sims typically open with an initial interaction that focuses students' attention (Pea R. D., 2004), perceived through color, object placement, object design and other cueing on the screen. We rarely use text to encourage an initial interaction. This initial interaction provides intrinsic motivation (Lepper & Chabay, 1985) for productive sim usage in several ways: to indicate to the student that the sim requires interaction, not passive observation; to engage the student in asking questions, typically around the resulting feedback from their interaction; and to provide the student with an immediate positive experience, so they know they can successfully use the sim and gain confidence as sim explorers (Scardamalia & Bereiter, 1991; Moos & Azevedo, 2008).

These goals for students' initial interaction and their framing of sim use are accomplished through design, by ensuring that the initial interaction is obvious, pedagogically relevant and results in feedback for the student. Use of familiar representations cues students to productive interactions – allowing students to effectively perceive the affordances of the sim. Representations and actions that draw on intuitive knowledge build on students' life experiences to ensure that they accurately interpret the resulting feedback. For instance, we design with familiar controls (buttons, sliders, etc.) that cue students to their interaction mode (e.g., buttons can be pushed) (Norman D. A., 2002). The representations in the sims are cartoon-like and colorful, reflecting a playful and inviting learning environment, rather than a technical-looking environment that might require instructions. Interactions are also low-risk, big, typically brightly-colored 'reset all' button or icon sets an entire sim tab to its initial state, and many actions can be undone quickly and easily (e.g., moving a slider back to its initial position).

*Scaffolding to Enable Sense Making*

When students begin attempting to answer their questions through interaction with the sim, we want the sim design to enable students to engage in productive inquiry – where their attempts at sense making with the sim generate constructive questions and ideas, followed by further exploration for answers. To do this, we design the overall layout, representations, feedback, range of interactions and illuminating cases to support productive sense making.

**Representations:** Each sim consists of representations for students to interact with, e.g., objects to move, charts and readouts that change based on student interaction with objects, controls to vary and buttons to push. These representations mediate how students construct understanding with the sim (Podolefsky, Perkins, & Adams, 2009). Representations in each sim are designed to be familiar and to draw on intuitive knowledge. If a representation is not familiar, a goal is for it to become familiar, to allow students to construct knowledge about this new representation through exploring its relationship to other, more familiar, representations. Representations change dynamically, based on the interactions of the students, maintaining students as the key agent of the learning experience. Making sense of the relationships between representations is often key to students understanding the concepts addressed in the sim. (Ainsworth, 1999; Berthold & Renkl, 2009) These relationships are indicated visually, often by proximity (labels are close to the object they label), by type or timing of feedback given (objects that give the same type of feedback – e.g., graph readouts or motion – are likely to be related), or by color (a control and arrow with the same color share a relationship). Some representations are chosen for their ability to do cognitive work for the student, highlighting aspects of a concept or making relationships salient. In many cases, canonical representations of a discipline provide an opportunity for students to construct understanding of expert representations.

**Feedback:** Feedback is a core element of scaffolding (Pea R. D., 2004; Quintana, et al., 2004). In simulations, interaction with the representations results in feedback – e.g., changes to motion, color, size, and readout values. This feedback serves multiple purposes. It provides students with immediate feedback that they have executed an action. An absence of feedback can be interpreted as "no action has occurred", cueing the student to try that action again, if necessary. Well-designed feedback is foundational to effective interface design. It can be distracting for a user to be uncertain if an action was 'registered' by a computer or not, so feedback that is readily interpretable and immediate allows for a seamless interaction for the student, maintaining a mode focused on conceptual understanding rather than sim usage. (Norman D. A., 2002) The appearance of feedback provides students an obvious point to begin asking questions, e.g., "Why did that happen?", "Does the same thing happen when I do this…?" Actions are repeatable, and, with changes in parameters, feedback can be variable, highlighting important concepts and causal relationships.

**Layout:** We build on several layout design principles from Mayer (2009). The overall layout of each tab is designed to encourage students to first explore the most foundational aspects of the sim tab before continuing on to the more advanced options available. We design the sim layout to scaffold learning of complex ideas without being overwhelming (De Jong & Van Jooligen, 1998). Students typically try to interact with colorful, centrally-located objects first, before exploring toolboxes and other peripherally-located controls. Students tend to explore toolboxes from the top to bottom. Putting the most foundational options for conceptual understanding at the top of toolboxes typically leads to students finding and exploring these options early in sim use. Options lower in the toolbox are typically found and explored later in sim use. We also cue students to recognize relationships between objects or controls by grouping them together.

We minimize wording used in the sims by placing short labels near objects. Minimizing wording can result in student's sense making about the words, rather than looking to the words as instructions. When students are unfamiliar with a word in a sim, they can use interactions with the representation(s) it refers to in the construction of an operational definition of what the word means.

**Range of interactions:** The range of possible student interactions is carefully chosen, so that student exploration is afforded flexibility but bounded within a parameter space that is productive for learning. (Pea R. D., 2004; Podolefsky, Adams, Lancaster, & Perkins, 2010) We do this by constraining the area of the screen in which objects can be moved or occupy, e.g., most representations can only be moved within a sim's 'play area', and cannot be placed, for example, inside a toolbox. We also place constraints on the conceptual parameter space that students are able to explore, through selection of the boundaries of variable parameters (e.g., the value range of sliders, the fastest or slowest an object can move, the most or least amount that can be added).

Being able to set these boundary conditions is both a productive constraint and a significant affordance of sims. From observations in interviews and classrooms, we find that students explore the extreme conditions. Through design, we can productively tap into this tendency by selecting boundary cases to highlight useful comparisons for students. If a student, for example, compares what happens when friction is at its highest versus when there is no friction, we can use those scenarios to clearly highlight aspects of a concept for students. To aid in making concepts salient through comparisons and contrasts, we also design in easy-to-interpret feedback, e.g., sliders and readouts that support exploring simple ratio relationships (e.g., twice as much, or half as much).

**Specific Illuminating Cases:** The features discussed above – e.g. representations, layout, feedback, etc. – are designed in coordination so that learners are likely to encounter illuminating cases through exploration – that is, encounter sets of pedagogically useful scenarios or cases that help learners recognize key relationships and make sense of concepts (Redish, 1994). However, for some content areas, we use unique design features to enable students' access to specific illuminating cases that serve as particularly useful scenarios for comparisons. For instance, use of a dropdown menu with careful selection of the options can make particular illuminating comparisons readily available. Within a tab, scene-selection buttons can be used to provide shortcuts to particularly relevant sim conditions and encourage exploration of those sim conditions.

*Scaffolding for Continued Engagement and Sense Making*

Once students have explored a sim for some time, we want the sim design to encourage continued engagement and sense making. To achieve this, we design puzzles and challenges into the sims.

**Puzzles and Challenges:** Conceptual puzzles are embedded through features or situations within the sims that students can use to challenge themselves and can figure out through sim use – for instance, interpreting a sim feature, or making a particular action occur that requires conceptual understanding. Each puzzle is intrinsic to the sim context. Challenges are often implicit goals present in the sim that students can work towards, such as making sense of a behavior or representation in the sim. Challenges are sometimes explicit, such as questions that encourage students to explore aspects of the sim for specific conceptual understanding. These explicit challenges do not direct students in how to use sim exploration to correctly answer a question. Rather, they provide a question that promotes productive sim exploration – students must determine what understanding they need and how to obtain it from the sim. Note that the sim we describe below includes implicit challenges rather than explicit questions or tasks for the student.

These puzzles and challenges provide students with motivation to explore and opportunities for built-in self-assessment – opportunities for students to identify inconsistencies in their knowledge or to determine if they have learned the concepts the sim focuses on.(Melero, Hernandez-Leo, & Blat, 2011; Tabak, 2004)

*Summary*

These strategies for implicit scaffolding in sims – scaffolding of the general concept, of students' framing of sim use, to enable sense making, and for continued engagement – work together to enable greater student agency and ownership over their learning, while both affording and constraining actions that are productive for learning. Students perceive the sims as engaging, open exploration spaces, while implicit scaffolding provides cuing and guidance so students are inclined to interact with the sim in productive ways. In other words, implicit scaffolding guides without students feeling guided.

These strategies are built upon models of learning, tool-mediated learning, and tool design, and are influenced by prior research on educational design. Each scaffolding strategy supports students' ability to construct their own understanding of the concepts embedded in the sim. Through careful design, each sim mediates the ways in which students engage with the sim content, leveraging affordances and constraints to ensure effective tool use.

# Example Application to Simulation Design: *Energy Skate Park: Basics*

Here, we show an example of the implicit scaffolding designed into the *Energy Skate Park: Basics* (*ESPB*) sim. Using the implicit scaffolding framework, we first describe the sim's features. We then present a case study of a student's use of *ESPB*, highlighting the role of implicit scaffolding.

*Scaffolding the General Concept*

The scope of the *ESPB* sim includes basic concepts of energy exchange. The specific learning goals chosen for the sim include:

1. Explain conservation of energy using kinetic and gravitational potential energy.
2. Describe how kinetic and gravitational potential energy relate to speed and position.
3. Explain how changing the skater's mass affects energy.
4. Explain how friction affects the skater's energy and motion.

The sim is sequenced through the use of three tabs, *Introduction*, *Friction* and *Track Playground*, shown in Figure 3. Each tab allows students to select or build tracks for a skater. This general setting provides students with the opportunity to explore a familiar scenario (skateboarding and tracks) to interact with an object above the ground, on the ground and in motion – components conducive to understanding kinetic and gravitational potential energy (which we refer to as "potential energy" in the sim). Each tab provides additional interactions and representations that support individual learning goals.

In the *Introduction* tab's upper left corner, students can select from three pre-set tracks. In the toolbox in the upper right corner, students have access to energy representations that change dynamically as the skater moves – an energy Bar Graph and Pie Chart that show relative amounts of kinetic, potential and thermal energy, as well as the total energy. Students also have access to: a Speed meter, which provides a qualitative measure of the skater's speed; a Grid, which overlays the tab with a grid allowing for students to accurately reproduce motion scenarios; and the "Skater Mass" slider, which allows students to change the mass of the skater.

The *Introduction* tab supports Learning Goals 1, 2 and 3. As students interact with the tracks and skater, they can observe the dynamic energy representations available through the Bar Graph and Pie Chart, which supports understanding of conservation of energy through the exchange of kinetic and potential energy (Learning Goal 1). Students can also utilize the Speed meter and Grid to explore relationships between energy forms, speed and position (Learning Goal 2). Through exploration of the "Skater Mass" slider, students can determine how mass affects energy (Learning Goal 3).

Several pedagogically appropriate simplifications implicitly scaffold student's interaction in the *Introduction* tab. In this tab, students are constrained to explore the exchange between kinetic and potential energy, and the relationship to total energy, by simplifying the system such that thermal energy is zero. To do this, friction is set to zero and cannot be changed on this tab. Also, the skater "snaps" to the track if released near the track, and "sticks" to the track while in motion – not allowing for bouncing or jumping off the track that results in changes in thermal energy. Note that in other tabs, this scaffolding is faded as "Stick to Track" can be turned off and friction can be increased.

The *Friction* tab contains the same features as the *Introduction* tab, with the following differences. Instead of having a "Skater Mass" slider, the skater's mass is fixed, and a "Friction" slider can be interacted with. This feature productively constrains student exploration to the effects of friction rather than mass (Learning Goal 4). Friction can also be turned "On" or "Off" with radio buttons. Additionally, there is the option to turn "Stick to Track" off, and explore the effect of bouncing and jumping on energy conversion between forms.

The *Track Playground* tab contains all the same features as the *Friction* tab, with relaxed track building constraints as compared to the *Introduction* and *Friction* tabs. This tab allows students to build custom tracks to create more complex scenarios during energy exploration.

*Scaffolding for Framing of Sim Use*

The state of the sim when first opened is carefully designed to frame the students' early sim experience as one of interaction and exploration. *ESPB* opens to the *Introduction* tab, which contains representations familiar to most students, a skater and a skate ramp, or "track". The design is cartoon-like and colorful, meant to create an inviting and unintimidating interface. The skater is stationary on the ground, requiring action and providing an invitation for the student to immediately interact with the sim. The track initially selected is a simple U-shape, inviting students to drop the skater "inside" the track, making the initial interaction obvious and intuitive. Since the skater stays inside of the U-shaped track, the students' first interactions with the sim are interesting and easy to understand.

Track selection buttons in the upper left of the tab are used to select between the three track options – the U-shaped, Ramp, and W-shaped tracks, shown in Figure 4. These large, colorful buttons invite interaction and exploration of the three tracks. The track shapes are implicit cues for students to explore how the skater's energy changes with speed and position. For example, the U-shaped track invites exploration of kinetic and potential energy that exchange repeatedly as the skater moves back and forth along the track, while the Ramp invites exploration of constant KE (and zero PE) after the skater rolls away from the bottom.

All of the readouts in the toolbox are off or set to a moderate position (Bar Graph, Pie Chart, and Speed meter are initially closed, the Grid is not selected and the skater's mass is set to a moderate point) in order to: 1) keep learners from being overwhelmed with visual information, and 2) to induce learners to focus on readouts as they are opened. Intuitive controls – check boxes – are used to turn features on and off. In this way, students' can figure out how to change the visual information available by simply "clicking around". If the visual information is at any time overwhelming or distracting a student, the student can easily turn items off to focus and then back on to self-adjust complexity as needed. Feature labels use minimal wording, generally one or two words, encouraging framing of the activity as one of exploration through interaction, rather than extensive reading. These design choices lead students to explore the sim by implicitly suggesting that the sim is a low-risk environment where they can "mess about" (Hawkins, 1974), explore, and learn from the sim without breaking anything or doing something wrong. Students can quickly reset the tab back to its starting conditions by clicking on the brightly colored "Reset All" button beneath the toolbox.

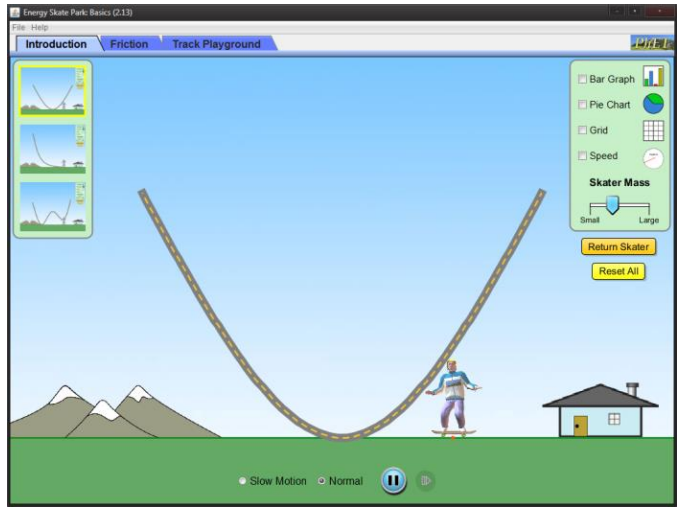

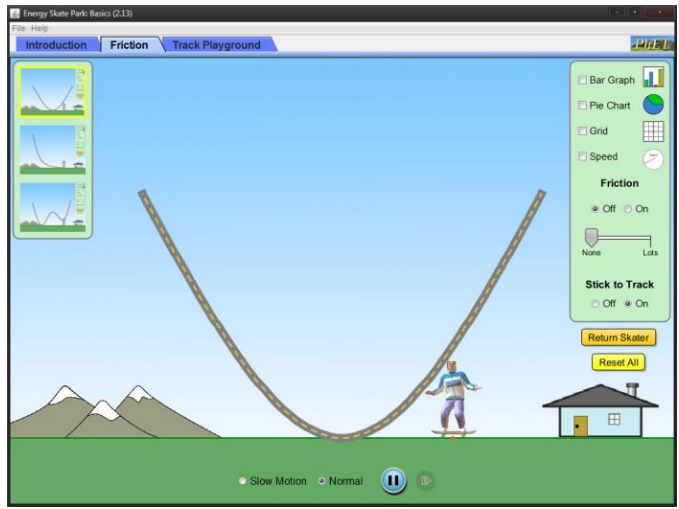

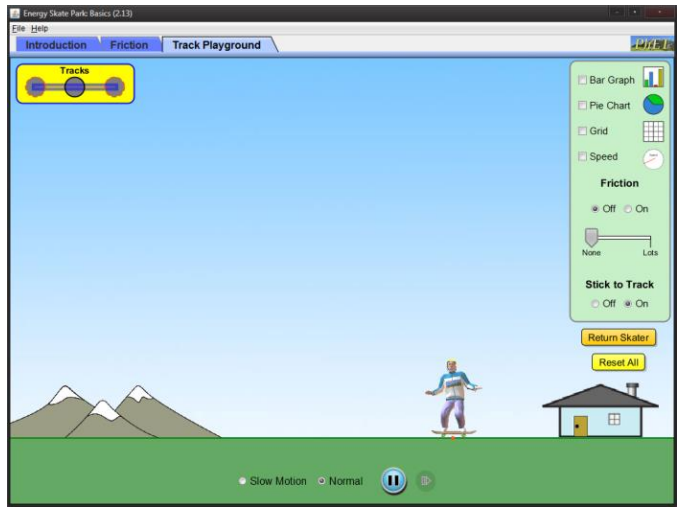

**Fig. 3** *Energy Skate Park Basics* tabs, from top to bottom: *Introduction* (starting configuration), *Friction,* and *Track Playground*

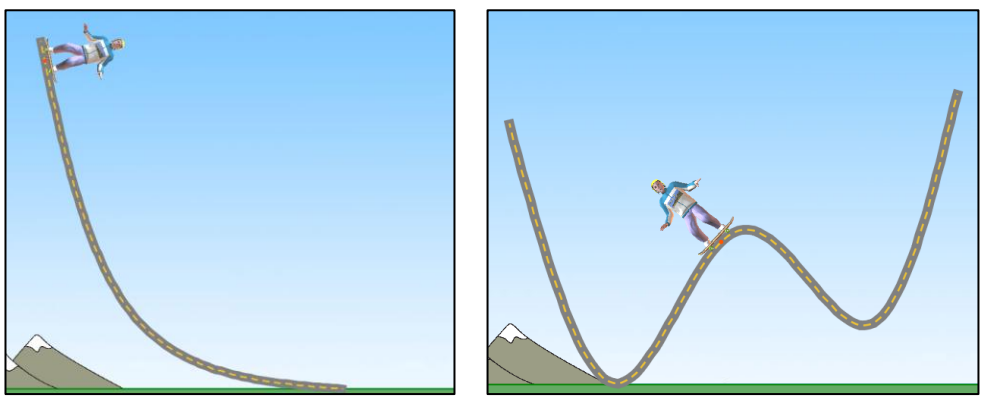

**Fig. 4** Images of the *Ramp* (left) and *W-shaped* (right) tracks, taken from the play area

*Scaffolding to Enable Sense Making*

The representations function in real time and in coordination with each other. Two energy representations are provided for students (Bar Graph and Pie Chart). These representations draw on student intuitions, such as "higher is more" in the Bar Graph, or "larger is more" in the Pie Chart. If one representation is more familiar they can use their knowledge and exploration to build an understanding of the other. By building on these intuitions for meaning making, students can begin to recognize relationships on their own. The representations in the sim cue students to make sense of the readouts, e.g., presence of the Speed meter cue students to track the skater's speed at various locations on a track. A legend indicates the color-coding for kinetic, potential, and thermal energy.

The feedback from each representation changes in real time. For example, in the *Introduction* tab, using the "Skater Mass" slider results in real time changes to the skater's mass, indicated by a change in size of the skater. Increasing the skater's mass results in an increase in kinetic and potential energy, which is visible in the Bar Graph or Pie Chart immediately as students move the mass slider. In the *Friction* tab "Friction" can be turned on or off with radio buttons, and adjusted from "None" to "Lots" with a slider, allowing students to explore the effect of friction on thermal energy. Since total energy is conserved, increasing thermal energy means kinetic and potential energy are reduced, indicated by the Bar Graph and Pie Chart readouts.

Illuminating cases scaffold student sense making of energy concepts. The three tracks chosen for the *Introduction* and *Friction* tabs target specific illuminating cases useful for the sim's learning goals. The U-shaped track provides the illuminating case where students can observe the skater's kinetic and potential energy exchange completely as the skater moves from the top to bottom (while total energy remains constant). The skater will move back and forth in this track repeatedly without leaving the track, so students have ample time to examine the energy exchange afforded by the U-shaped track. The Ramp track is a single-sided ramp, providing the illuminating case where students can observe the potential energy convert to kinetic energy, while the kinetic energy remains constant as the skater moves along the ground. The W-shaped track provides the illuminating case where students can observe kinetic and potential energy exchange in a more complex way than the U-shape.

We leverage the layout of the track buttons in the *Introduction* and *Friction* tabs so that if students select tracks from top to bottom, which is typically the case, they interact with the tracks based on increasingly complex ideas. The use of large, attractive buttons affords quick and intuitive access to different track configurations, with a pedagogically useful increase in complexity. The pre-defined tracks also constrain students to only these track configurations, which have been carefully chosen to promote exploration of illuminating cases. We introduce the affordance of custom track building in the *Track Playground* tab, so that students have typically explored energy under the more constrained *Introduction* and *Friction* tabs first. However, track building still has some constraints in order to promote productive exploration – a maximum of four track pieces are available, effectively constraining the size of the track that students can construct.

Floating toolboxes allow us to divide controls into logical groups; helping students more easily parse the sim interface. We place commonly used controls, such as "Return Skater" and "Reset All" in bright buttons separated from the control panels, making these controls highly visible. The Pie Chart remains directly above the skater, using proximity to cue the relationship between the chart and the skater's motion. No features are hidden or buried in menus – they are all accessible immediately and result in immediate feedback when used. Students can quickly find the features and determine what each does as they "click around".

In all tabs, the "Return Skater" button puts the skater back in the last dropped location. This supports students' experimentation by allowing them to repeat the same action multiple times, or make slight changes over multiple trials. It will also bring the skater back if it moves off the screen. The sim can be set to "Slow Motion" or "Normal" speed using radio buttons at the bottom of the tabs. "Slow Motion" allow students to carefully examine changes in energy as the skater moves. Finally, the sim can be paused using the "Play/Pause" buttons, freezing the action on the screen. While paused, the sim can be "Stepped Forward" in small time increments.

*Scaffolding for Continued Engagement and Sense Making*

Typical with all PhET sims, students initially engage in sense making through interactions with the immediately obvious representations (e.g., the skater and track), and then explore the variety of tools available. Students then typically continue their sim use through the second and third tabs. In *ESPB*, the final tab is the *Track Playgound* tab, which supports students in creating their own challenges by allowing students to build custom tracks (See Figure 5). The *Track Playground* tab starts with an empty play area, cueing students that they need to drag track pieces from the "Tracks" box. Track ends "snap" together, and the tracks can be bent into a variety of configurations.

The *Track Playground* tab is typically reached later in student use, after students have explored the *Introduction* and *Friction* tabs, and have begun asking questions appropriate for more complicated track designs. For example, once students relate friction to thermal energy, they may try to determine where the energy goes when the skater slows to a stop and if the outcome changes based on track configuration. In the *track playground*, it is common for students to challenge themselves to make loops in which the skater will skate upside down, as shown in Figure 5. Students can self-assess their understanding by making changes and checking results against their predictions, for example: creating experiments to check whether energy is conserved under different conditions, moving the skater up and down to see how height affects potential energy, or watching the Speed meter and kinetic energy bar change simultaneously.

Note that in *Energy Skate Park: Basics*, there are no hard constraints on when students can use any given feature. In some cases, a student may click to the *Track Playground* tab early in sim use. In this case, students can explore this tab, and as questions arise they can explore the previous tabs for more focused investigation (e.g., to explore how mass affects the skaters speed on a track). We believe that this flexibility supports student agency, allowing students to take control over their own learning path. With the use of implicit scaffolding, the sim simultaneously provides guidance so that students tend to explore in a productive sequence.

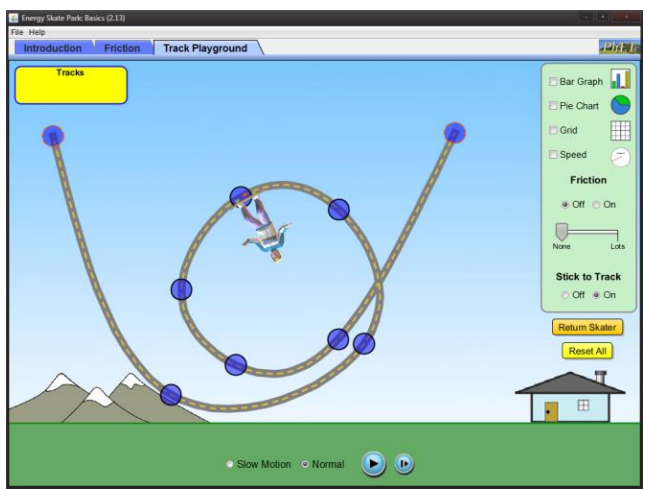

**Fig. 5** ESPB Track Playground

# Empirical Support

*Methods*

To assist in the development of an effective PhET sim, we conduct individual student interviews on every sim developed. For *ESPB*, nine middle school students were interviewed during the design process. For the purposes of this work, we present results from one of these student interviews, conducted on the final version of the sim – available at http://phet.colorado.edu/en/simulation/energy-skate-park-basics. This student was female and in 6th grade.

During all of the interviews, students were asked to 'think aloud' while 'playing' with the sim. These students had not used the *ESPB* sim prior to the interview. While a student is using the sim, prompts from the interviewer are only used to encourage the student to talk out loud. These prompts are unscripted, open-ended questions, e.g., "What do you think is happening?" or "What are you trying now?". Other than these prompts, the interviewer is quiet through most of the interview. The interviews are video recorded with a camera, and screen capture and audio recorded using the software Camtasia (Techsmith).

The purpose of this interview data is to ground the theoretical framework described empirically, and illuminate how it can apply to a student using a sim including implicit scaffolding. We do not claim that all students will engage in the way we describe here. This would require additional study with a different experimental design that is beyond the scope of this paper (but is the subject of current and future work). Nonetheless, the engagement and learning in this interview is similar to other interviews conducted using *ESPB*, as well interview findings with students using other PhET sims. (Adams, et al., 2008) While limited in what can be generalized, the strength of this dataset is its richness and detail. (Otero & Harlow, 2009; Hatch, 2002) The micro-analysis presented here draws on a dense qualitative dataset, and provides insights into how implicit scaffolding works in practice.

*Interview Data*

Here we provide an example of effective sim use supported by implicit scaffolding designed into an interactive sim. In this case study, a 6[th] grade girl uses *ESPB* without explicit external guidance (neither written nor verbal). While classroom use of a sim will generally include some external guidance, here we focus on student sim usage when supported only by implicit scaffolding. Qualitative microanalysis of this interview data highlights the usefulness of implicit scaffolding for supporting student engagement and learning.

In the transcript below we include *all* of the student's verbalizations (in quotes) during the first 5 minutes and 30 seconds of using the sim. In sim interviews, substantial insight into student thinking is gained through analysis of sim interactions. Therefore, in addition to verbalizations, we include detailed descriptions of the student's sim interactions (in italics) in the transcripts below. During this time the interviewer observes without speaking. Time is indicated in the mm:ss format, starting from the moment the sim opens on the computer screen.

**Excerpt 1: Student's first interactions with the sim.**

- 0:02 [*Sim opens with U-track selected and skater stationary next to track. Student makes sim fullscreen.*]
- 0:06 [*Clicks U-track button, then Ramp track button. Track changes from U-track to Ramp track.*]
- 0:10 [*Clicks and drags skater to top of Ramp track*] Student: "Interesting." [*Drops skater onto track, skater rolls down track and off screen.*]
- 0:24 [*Selects W-track, picks up skater and drops on left side of track. Skater rolls to the right side of the track then back towards the left side of track. Student clicks pause button.*]
- 0:44 Student: "I guess it's kinda interesting how, like, once he goes down, like, that movement, like, transfers into the uphill. Almost like the gravity, like, almost, it almost seems like it, like, transfers once he goes back up."

Notice how quickly the student began interacting – in less than 10 seconds – and trying to make sense of what was happening in the sim. The design of the sim, the starting representations and display of tools, did not inhibit her inclination or ability to explore immediately upon the sim opening. The student's first interaction was to change the tracks, starting from top to bottom, and then to drop the skater on her chosen track. The design of the sim supports the ease of these interactions; selecting one of three large track buttons results in track changes and putting the skater on the track requires only click-and-drag interaction. This ease of use appears to support her focus on sense making with the sim, as demonstrated in the last line.

Within the first minute she verbalizes her attention to the motion of the skater and how the motion changes as the skater moves along the track, a key component of using the sim to learn about kinetic and potential energy exchange. The parameterization of the sim into interactions that result in changes in motion of the skater (through the use of different tracks) may have cued the student to attend particularly to skater motion.

The student not only attends to the skater's motion, she also verbalizes an idea to try to make sense of the behavior she is observing. She said this behavior might involve "transfer" of some type, indicating an attempt at explanation for the behavior. This type of explanatory verbalization signals that she is in a sense making mode, making careful observations of the behavior of the sim's representations, and beginning to formulate ideas about her observations. These ideas turn into questions she can test, leading to further exploration and sense making. In the following excerpt, she tests the effect of the skater's mass on motion.

**Excerpt 2: Effect of Skater Mass**

- 1:01 [*Moves "Skater Mass" slider to "Large", moves skater to top left of W-ramp. Clicks play button. Skater goes down ramp, then up middle bump.*]
- 1:14 Student: "He definitely seems to go up [the ramp] faster when he's bigger."

1:23   [*clicks pause button once skater has reached the right side of ramp, and has started back down to the left. Moves "Skater Mass" slider to "Small", clicks play button.*]

1:25   [*Skater moves over middle bump to the left and reaches the top left side.*]

1:31   Student: "Not as fast on the way down when he's small, and then—whoa, he just fell off [*skater falls off left side of track.*], but I'm not sure [laughs] if that's really part of it. Um."

Here the student uses the extremes of the "Skater Mass" slider to make qualitative interpretations of the effect of the skater's mass. The real-world appearance of the slider cues her to immediately use and try to make sense of this control. The mass of the skater changes dynamically, with feedback as the skater image increases or decreases in size as its mass changes, so she can quickly compare the motion of a "small" skater to the motion of a "large" skater. The playfulness of the skater being able to fall off the track is amusing, framing the sim as an environment for low-risk exploration, without distracting from her focus on sense making.

The student appears to determine that the speed of the skater depends on its mass. In fact, the skater's speed is independent of mass, and this correct behavior is shown in the sim. We believe that this student's perception may be due to a well-documented student intuition that heavy objects move faster (Hestenes, Wells, & Swackhammer, 1992; Kavanagh & Sneide, 2007; Pine, Messer, & St. John, 2001), prior knowledge she brings to the interview and not something indicated in the sim. However, later in the interview (at about 26 minutes) this student returns to the *Introduction* tab, unprompted, and performs further experiments to examine the relationship between mass and speed. She correctly concludes that both large and small skaters move at the same speed when dropped from the same location. The flexibility of the sim and the availability of tools, such as the speedometer, support this student in revisiting and revising her thinking.

She next moves on to explore the *Friction* tab.

**Excerpt 3: Initial Exploration of *Friction* tab and "Friction" slider.**

1:40   [*Student switches to "Friction" tab, which starts with U-track showing and skater stationary next to track.*]

1:42   [*Clicks and drags skater above track, drops skater onto track. Skater moves back and forth on track.*]

1:51   [*Student clicks on Bar Graph tool, Bar Graph opens next to U-track. The Bar Graph shows the exchange of kinetic and potential energy as the skater moves on the track, with the total energy remaining constant.*]

2:02   Student: "It's kind of interesting how the kinetic energy goes up when he's going down, and the, umm, potential goes up when he's going up."

2:12   Student: *"*Thermal [bar graph] for this one seems to be staying down most of the time."

In this excerpt, the student begins directly exploring kinetic and potential energy – Learning Goal 1 – after less than two minutes of sim interaction. After opening the Bar Graph and observing the skater's motion, she makes the connection between the motion of the skater and the exchange between the kinetic and potential energy (at time 2:02). Rather than requiring explicit instruction, many design features of the sim supported the student in making this connection. The U-track was pre-selected and readily available to provide the specific illuminating case of simple kinetic and potential energy exchange. The action of dropping the skater inside the track to begin motion cues that the motion of the skater is significant and centers exploration around relating skater motion to other representations in the sim. The pedagogically appropriate simplification of starting this tab with friction "Off" allows the skater to move back and forth continuously on the U-track, making comparisons with the Bar Graph easier than if the skater slowed to a stop and its motion had to be restarted.

The Bar Graph itself was designed to highlight useful energy comparisons and support sense-making around those comparisons, such as the observations made by this student at 2:02. The energy types are color-coded and clearly labeled, kinetic and potential energy bars are next to each other to cue comparisons and the graph is scaled so that the dynamic changes are clearly visible. While being an expert representation, the Bar Graph draws on student intuitions such as "up is more" (diSessa & Sherin, 1998). The real time feedback provided by the Bar Graph allows for quick connections to be made between the skater motion and the graph changes. The text on the graph is minimal, promoting sense making through coordination of the text, Bar Graph, and skater's motion.

As she continues exploring, she selects options from top to bottom in the toolbox. The placement of the checkbox to turn on the Bar Graph at the top of the toolbox cues her to "try this first", resulting in access to a useful representation to make sense of kinetic and potential energy exchange in conjunction with the U-track (the first pre-set track). She then moves on to the Pie Chart tool.

2:24   *[Clicks on Pie Chart tool, Pie Chart opens above the skater. Closes Bar Graph tool window.]* Student: "Umm." [Student clicks on Grid tool, Grid appears in sim.]
2:32   Student: "I think it's a little bit easier to see it [change in kinetic and potential energy] on the Pie Chart."
2:43   Student: "The Grid makes it a little bit confusing." *[Student clicks Grid off. Grid over sim disappears.]*
2:48   *[Student clicks on Speed meter tool, Speed meter tool appears above track. As the skater moves along the track, the Speed meter shows the change in the skater's speed.]*
3:00   *[Student clicks on "Friction" slider – does not move – then pauses the sim, clicks "Reset Skater" button. Skater moves to top left side of U-track. Student selects Friction "On" radio button, then moves "Friction" slider from "none" to "lots". She clicks "Return Skater" button again, then clicks play. Skater moves slowly down track, with Pie Chart showing increasing amounts of thermal energy.]*
3:22   Student: "He seems to go slower with the friction, especially on the way up [the side of the track] – he doesn't go all the way. Umm…until I *[moves "Friction" slider from "lots" to "none"]* return the skater [*clicks "Return Skater" button*] and put it ["Friction" slider] back to none."

The student recognizes that the Pie Chart and the Bar Graph are representing the same thing (energy types). Implicit scaffolding is employed to cue this coordination between representations through the color-coding used for both representations, the consistent labeling of energy types, and the synchronized, dynamic changes both representations show as the skater moves. Her use of the two representations and the comparison between them – using them to make sense of the relationship between energy exchanges and motion – is an example of the sim supporting sense making with expert-like representations.

She also readily perceived that she could make her own path through the sim, changing to the *Friction* tab when she chose to without asking for permission or expecting a prompt, and also selecting representations as she felt appropriate. She compared the Bar Graph and Pie Chart, and also turned on the Grid. She decided that the Grid was confusing at that point, and so she chose to turn it off and had no difficulty in doing so. Note that later in the interview she will turn the Grid on again, and use it for measuring distances. This highlights one of the advantages of using implicit scaffolding to support student-directed exploration – students often find features that are not used immediately, but prove useful later on. The fact that students "get to know" a sim through exploration allows for further sense making by using tools to answer questions as they arise for students.

Her exploration of the toolbar reaches the Speed meter and the "Friction" slider, and here she starts exploring the relationship between friction and speed – determining that more friction results in a slower skater. She continues this line of inquiry into the next tab.

**Excerpt 4: Initial use of Track Playground**

3:40  [*Student drags out track pieces from "Tracks" box, ultimately arranging three track pieces into asymmetric W-track shown in Figure 4.*]

4:16  [*Student drags skater above track, and drops onto track. Skater moves along track, jumping off the end.*] Student: "Whoops. I think I just killed him."

4:31  Student: "Umm, I guess it's kind of interesting, umm, what happens, like, with, like the first down got [*moves mouse cursor down left side of custom W-track*], like, faster and faster and then here [*moves mouse cursor from top of center bump along right side of custom W-track*], he was still maintaining the same speed on the small up to some extent."

4:49  [*Clicks on Bar Graph tool, Bar Graph opens next to track. Clicks "Return Skater" button, skater appears at top left of custom W-track and moves down track.*] Student: "He has a little bit of thermal energy on this one definitely more than the other courses I did." [*Skater jumps off end of track. Clicks friction "On" radio button, Friction "Slider" moves to center. Clicks "Return Skater" button, skater appears at top left of custom W-track and moves down track.*]

5:14  Student: "Umm. Thermal energy goes up again, with the friction it seems *[skater does not make it up center bump]*. But definitely he can't go through the course, umm, with the friction turned on as much [*moves "Friction" slider from center to just above 'none'*] until maybe I lower it." [*Skater moves over the center bump*]

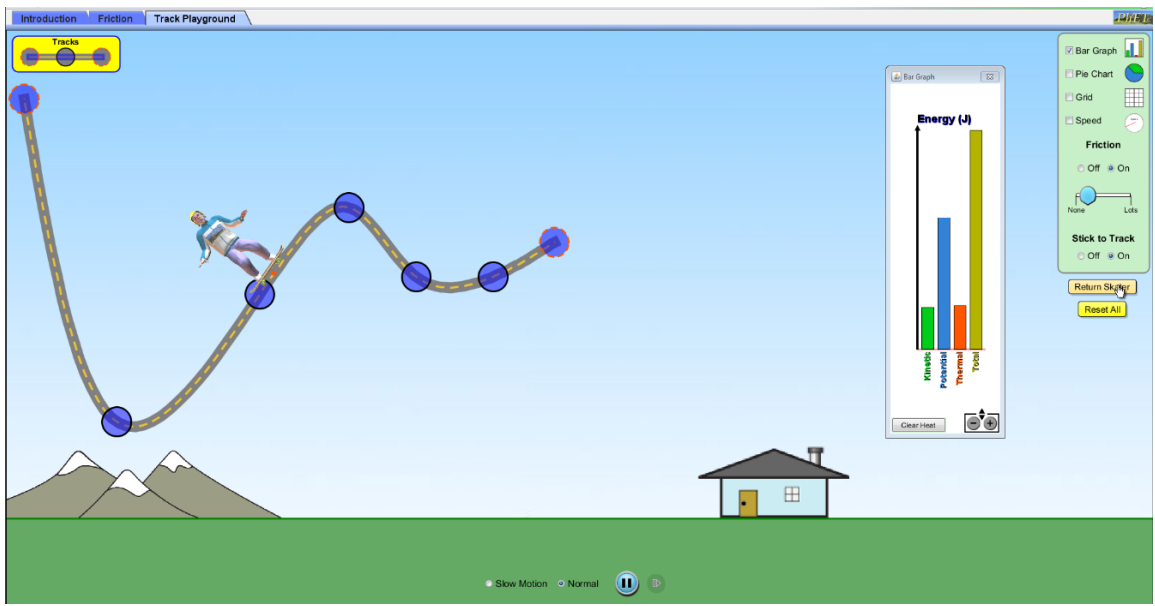

**Fig. 6.** Screenshot of student's asymmetric W-track

In this excerpt, the student utilizes the relaxed constraints of the *Track Playground* to build her own version of a W-shaped track (Figure 6). She then carries out experiments, changing the friction each time and making observations of the change in the skater's motion, ultimately concluding that thermal energy and friction are linearly related, and the ability of the skater to move along the entire track depends on the amount of friction present.

In these 5 minutes and 30 seconds of excerpts – the student's first minutes interacting with the *ESPB* sim – this student readily interacted with the sim and engaged in exploration and sense making around the various representations available and the ideas of kinetic and potential energy exchange, friction, and mass. During the remaining 25 minutes of the interview she continues to engage in experimentation and sense-making with the sim, as described in the list below.

**She explores the effect of friction and mass changes:**

- Determines exactly how much friction – in terms of location of the "Friction" slider - can be present and still have the skater complete the course and that increasing friction decreases the height of the skater's jump trajectory.
- Determines that thermal energy is "heat energy" and that this exists when there is friction, since "friction usually creates heat".
- Returns to the *Introduction* tab to explore the mass of the skater again, and uses the speedometer to determine that skaters with large and small mass move at the same speed when dropped from the same location.

**She explores the relationship between track configuration and skater motion:**

- Extends her initial asymmetric W-track to include a segment that touches the ground and then points up in the air. She uses this to determine how far uphill the skater can move after coming down a consistent length of downhill track.
- Explores the relationship between the steepness of a set length of downhill track and how far uphill the skater can move.
- Determines that the Grid can be used to measure the distance of the skater's motion. Uses the Grid to determine the relationship between the height of a downhill track and how high the skater can jump – measuring the highest point of the skaters jump trajectory. Explores effect of changing the height of the uphill portion of the jump.
- Analyzes the speed of the skater during each portion of its motion – downhill on track, uphill on track, up into the air and falling down to the ground.

**She explores the relationship between kinetic, potential and thermal energy:**

- Determines that track motion is not necessary for the skater to "have energy". Explores how clicking and dragging the skater up and down affects the energy types (using the Bar Graph representation). "When I raise him up and down he seems to have energy, which I didn't expect."
- Determines that at the top of a track there is a lot of potential energy, as the skater moves or falls towards the ground the kinetic energy increases, reaching a maximum at the ground. Once along the ground, with friction on the thermal energy increases as kinetic energy decreases.
- Relates movement of the skater on the track to kinetic energy, height of the skater in the air to potential energy.

Throughout the interview, she moves easily between the three tabs depending on her line of inquiry, and verbalizes predictions and observations in terms of energy types and motion of the skater.

# Discussion

As highlighted by Vygotsky (1978), the tools used in teaching and learning mediate how students engage with, perceive, and achieve the learning objectives (Figure 1). Sims that employ well-designed implicit scaffolding provide a significant opportunity to mediate students' interaction with science learning objectives in new and powerful ways.

With the new opportunities presented by these sims, it is worth examining how implicitly-scaffolded sims might influence the long standing debate over the goals and methods of education described in the introduction. The form and degree of scaffolding within learning environments is a central feature in this debate; some researchers cite the need for very explicit and direct instruction to achieve content learning (e.g., Kirschner, Sweller, & Clark, 2006) and other researchers cite the need for more open, student-centered instruction that enables participation in the processes and nature of science (Hmelo-Silver, Duncan, & Chinn, 2007). This tension over the nature and degree of scaffolding is readily apparent, even across implementations of inquiry-based learning environments where student activities and teacher facilitation range from pure discovery to heavily guided.

In Table 1, we have characterized these two extremes of inquiry-based instruction across several dimensions which shape the overall experience of the students. On one side of this *inquiry spectrum*, we describe *discovery* learning, characterized by minimal to zero explicit guidance for students. On the other, we describe *heavily guided* learning, characterized by specific procedural directions.

**Table 1** Characteristics of inquiry spectrum extremes

| **Discovery** | | **Heavily Guided** |
|---|---|---|
| Engagement, messing about, question asking | **Process of Learning** | Procedure following, question answering |
| High, individualized | **Variability of Learning Process** | Low, homogenous |
| Student, peers | **Agent in Learning Process** | Experts, teacher |
| Creator, explorer | **Role of Student** | Follower, (re-)constructor |
| Guide, co-participant | **Role of Teacher** | Provider, director |
| Evolving | **Nature of Knowledge** | Static, pre-determined |

These extremes differ significantly in terms of the goals and roles for participants (i.e., students and teachers). Discovery emphasizes student agency in the learning process and participation in the scientific discipline. Students take a leading role creating 'new-to-them' knowledge through active engagement with the resources, problems, or tools within the learning environment. The students engage as participants in doing science, while the teachers provide scaffolding through their role as a guide or co-participant in this process. However, discovery learning environments can present significant pitfalls, such as students not knowing what to do or following unproductive learning trajectories (Kirschner, Sweller, & Clark, 2006). In other words, the learning process and outcomes are often highly variable and individualized, which can lead to inefficiencies in achieving both process and content learning objectives and may have affective issues of students feeling confused or frustrated.

Heavily guided learning environments seek to reduce these risks of inefficiencies and poor or variable learning outcomes. By providing students with explicit scaffolding in the form of specific procedures or narrowly targeted questions, these environments focus on efficiently learning a well-defined domain of knowledge. Teachers direct the learning experience and serve as a key source of the knowledge that students are expected to learn. Increased certainty in the efficient learning of specific content is prioritized over the goals of student agency and student participation in the process of science.

We seek a middle-ground that achieves the productive aspects of both discovery and heavily guided inquiry (Fay & Mayer, 1994; Schwartz, Lindgren, & Lewis, 2009). We suggest that the use of implicit scaffolding within learning tools – specifically sims – provides a new option. In this option, the sim, through its design, supports student agency and participation while reducing or eliminating the issues of highly-variable learning outcomes and negative affective responses. Thus, we propose that by employing *implicit scaffolding* in the learning tool, we can achieve goals of discovery learning environments that are often missed by heavily-guided environments, while simultaneously constraining (somewhat covertly) the variability of learning.

As the interview excerpts demonstrate, an implicitly-scaffolded sim can support student agency – students' feeling of empowerment, that they can follow their own path and take ownership over their learning – while promoting exploration that is productive for student learning. Students are able to move freely through the sim – changing tabs, accessing options, resetting, etc. – choosing whether they want to explore a new area or explore again. Through careful design, each sim mediates the ways in which students engage with the sim content, cuing interactions and providing feedback that help students construct an understanding of the concepts embedded in the sim.

Thus, implicit scaffolding can change how students frame learning activities. The implicit, or indirect, nature of the scaffolding supports an alternative framing, wherein the answer to the question "what sort of activity is this?" is "an activity where I (the student) am in charge of my own learning". Our aim is to create an environment that encourages inquiry, is safe, and supports productive sense making while students explore along their own path. Within the structure of Table 1, implicit scaffolding influences *students' perception* of the inquiry approach along these characteristics. Students perceive themselves as engaging and messing about in open exploration spaces and creating knowledge, while implicit scaffolding provides tools, representations, and access to ideas that are designed to be useful to the student. In addition, cues and guidance lead students to interact with the sim in productive ways for learning. In other words, implicit scaffolding guides without students feeling guided.

Teachers as well as teacher-designed activities and worksheets continue to serve critical roles shaping and scaffolding the students' learning experience. We typically recommend that sims be used with worksheets and/or teacher facilitation, but we suggest that the design of the worksheet or the form of the facilitation can significantly leverage the presence of the implicit scaffolding towards achievement of these multiple goals. For instance, the presence of the implicit scaffolding provided by the sim allows activities to productively include a 5-10 minute period of open exploration, to use more open-style challenge questions to focus students' attention on specific learning goals, and to avoid explicit instructions of how to use the sim (Hensberry, Moore, Paul, Podolefsky, & Perkins, 2013; Moore, Herzog, & Perkins, 2013; Perkins, Moore, Podolefsky, Lancaster, & Denison, 2011; Podolefsky, Rehn, & Perkins, 2012). Teacher facilitation can solicit ideas from students – ideas students constructed from their interaction with the sim. Students overall learning experience is shaped by the dynamic interaction between the teacher, the students, the written worksheet (if used), and the sim. Implicit scaffolding built into the sim provides a unique opportunity to reimagine how the interactions and structures with this system can better support the multiple goals of education.

# Conclusion

The primary aim of this paper is to present a theoretical framework for simulation design, *implicit scaffolding*. The PhET Interactive Simulation project designs tools aimed at achieving multiple simultaneous learning goals, including process, participation, and content. To these ends, we have developed a framework for implicit scaffolding with the aim of guiding students without them feeling guided. This framework is built upon a foundation consisting of constructivist learning, tool-mediated learning and tool design, particularly in the context of human computer interaction. This foundation provides insight into the design of tools which employ implicit scaffolding to support construction of knowledge through exploration while allowing for student agency that may not be realized with more explicit guidance. Affordances, constraints, cuing, and feedback are key elements to achieving effective implicit scaffolding within learning tools.

We have demonstrated the application of implicit scaffolding in the design of a computer simulation for learning about energy. Qualitative data from a student interview grounds the theoretical framework, and serves as an example of implicit scaffolding at work. The strength of this data is its richness and detail – future work would use experimental designs meant to better establish the generality of our findings.

The implicit scaffolding framework is structured in terms of strategies for effective sim design. We have argued that implementation of these strategies results in sims that can achieve multiple educational goals simultaneously, and demonstrated the effectiveness of implicit scaffolding for promoting productive exploration by students. While our focus here is on computer simulations, the framework for implicit scaffolding is built upon a more general theoretical foundation. This framework may therefore be generalizable, and could likely be adapted for the design of other learning tools and environments, beyond computer simulations.

# Acknowledgements

We thank the entire PhET group and Physics Education Research group at the University of Colorado Boulder, and particularly Sam Reid for development of Energy Skate Park: Basics and Ariel Paul for conducting student interviews. We also thank TechSmith for contributing Camtasia software. This work was supported by the National Science Foundation (DRK12 #1020362), the Hewlett Foundation, and the O'Donnell Foundation.# References

Adams, W. K., Reid, S. R., LeMaster, R., McKagan, S. B., Perkins, K. K., Dubson, M. D., & Wieman, C. E. (2008). A Study of Educational Simulations Part I - Engagement and Learning. *Journal of Interactive Learning Research, 19*(3), 397-419.

Ainsworth, S. (1999). The functions of multiple representations. *Computers and Education, 33*(2), 131-152.25